\documentclass[11pt,twoside]{article}

\usepackage{asp2006}
\usepackage{epsf}
\usepackage{psfig}
\usepackage{lscape}

\begin{document}
\title{Outstanding Issues in Our Understanding of L, T, and Y Dwarfs}   
\author{J.\ Davy Kirkpatrick}   
\affil{Infrared Processing and Analysis Center, California Institute of Technology}    

\begin{abstract} 

Since the discovery of the first L dwarf 19 years ago and the discovery of
the first T dwarf 7 years after that, we have amassed a large list of these objects,
now numbering almost six hundred. Despite making headway in understanding the physical
chemistry of their atmospheres, some important issues remain unexplained. Three of these are the subject of this paper:

(1) What is the role of ``second parameters'' such as gravity and metallicity in shaping the emergent
spectra of L and T dwarfs? Can we establish a robust classification scheme
so that objects with unusual values of log(g) or [M/H], unusual dust content, or 
unresolved binarity are easily recognized?

(2) Which physical processes drive the unusual behavior at the L/T
transition? Which observations can be obtained to better confine the problem?

(3) What will objects cooler than T8 look like? How will we know a Y dwarf when we
first observe one?

\end{abstract}



\section{Introduction}

In the last thirteen years, the sample of L and T dwarfs has grown from a paltry 1 (GD 165B, \citealt{becklin1988}) to an astounding 586 at the time of this writing (see http://www.DwarfArchives.org). As the sample has grown, additional follow-up has further added to our knowledge of these objects.  With this newfound knowledge major deficiencies have begun to show in our understanding of the physics shaping the spectra of L and T dwarfs. In this paper I discuss three important questions being asked by low mass star and brown dwarf researchers today. First, what role do ``secondary'' parameters such as gravity, metallicity, dust content, and unresolved binarity play in shaping emergent spectra? Second, why has the transition between L dwarfs and T dwarfs been so difficult to model? Third, what will objects cooler than type T8 look like?

\section{Issue \#1: Gravity, Metallicity, and Other ``Secondary Parameter'' Effects}

\begin{figure}[!ht]
\plotfiddle{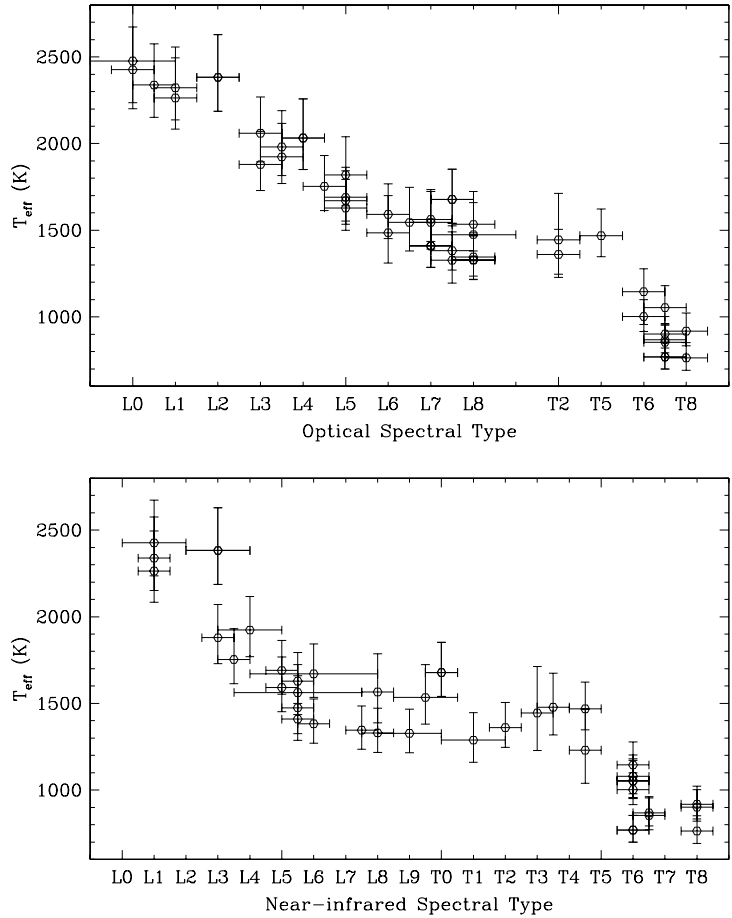}{4.7in}{0}{50}{50}{-145}{-20}
\caption{Effective temperature plotted against spectral type. 
The top panel shows the relation between temperature and optical 
spectral type; the bottom panel shows temperature versus near-infrared type.
See \cite{kirkpatrick2005} for details. 
\label{temp_type}}
\end{figure}

Astronomers have come to expect that spectral type can be used as a proxy for temperature. While this is true for main sequence stars classified from types O through M, is the same true for L and T dwarfs? Using measurements of trigonometric parallaxes (see \citealt{dahn2002, tinney2003, vrba2004}) and bolometric luminosities (see \citealt{reid1999, golimowski2004}), it can be shown that there is an amazingly linear correlation between {\it optical} spectral type and effective temperature for the entire L dwarf class. Effective temperature appears constant from early- through mid-T, then again shows a drop as a function of type for later T dwarfs. This is illustrated in the upper panel of Figure~\ref{temp_type}.

The same correlation, however, is not seen for {\it near-infrared} L and T dwarf classifications. This is illustrated in the lower panel of Figure~\ref{temp_type}.  Here the effective temperatures are essentially constant, albeit with a large scatter of ${\pm}$200 K, from mid-L through mid-T types. This is somewhat surprising given that the near-infrared classification is loosely based on the optical classification (\citealt{reid2001, testi2001, geballe2002}). However, these systems used only spectral indices for classification and not anchor points (primary spectral standards) to serve as on-sky comparisons. This differs from the methodology used in the optical, where anchor points were selected and checked for self consistency across the entire wavelength region used for classification, not just over the areas used for measuring spectral indices (\citealt{kirkpatrick1999}). In all fairness, the optical behavior of L dwarfs is much less complicated than the near-infrared behavior, as evidenced by the sometimes non-monotnic behavior of $H$- and $K_s$-band fluxes and shapes as a function of type (e.g., Fig.\ 3 of \citealt{geballe2002} and Fig.\ 10 of \citealt{mclean2003}), which is further reflected in the large spectral type error bars seen for some objects in the lower panel of Figure~\ref{temp_type}. Until very recently, setting up near-infrared L-dwarf spectral standards would have been difficult because of this scatter and the relatively sparse sampling of 1.0-2.5 $\mu$m spectra for each L dwarf subclass. 

What causes the large dispersion of morphologies seen in the near-infrared? As the optical spectra of L and T dwarfs do not show this degree of variation, we can use them to probe the influence of other parameters such as gravity, metallicity, and dust on the spectral shape at longer wavelengths.

\subsection{Exploring the Effects}

\subsubsection{Low Gravity}

The effects of lower gravity have been well studied in the optical for late-M dwarfs. The earliest studies of gravity-dependent features were based on young objects in the Pleiades (\citealt{steele1995,martin1996}) and $\rho$ Ophiuchi (\citealt{luhman1997}). At these ages (in the 1-100 Myr range) late-M dwarfs are substellar and are still contracting to their final radii; this means that they are both larger and less massive than old stars of the same effective temperature and hence will have lower surface gravity. Gravity-dependent features in the optical spectra of M dwarfs are well known as these are also the features used to distinguish luminosity classes (M dwarfs vs.\ M giants) on the MK classification scheme. In the far red portion of the spectrum, lower gravity results in weaker alkali lines and weaker hydride bands. At later M types, TiO and VO bandstrengths are stronger at lower gravity as well, and this occurs because of the gravity (pressure) dependent nature of condensation (\citealt{lodders2002}).

The study of low-gravity effects in L dwarfs, on the other hand, has had a brief history because few young cluster brown dwarfs with L spectral types are known. Recent discoveries of field L dwarfs believed to be young, low-gravity brown dwarfs have, however, recently been reported (\citealt{kirkpatrick2006, cruz2007}; Cruz et al., this volume).

\begin{figure}[!ht]
\plotfiddle{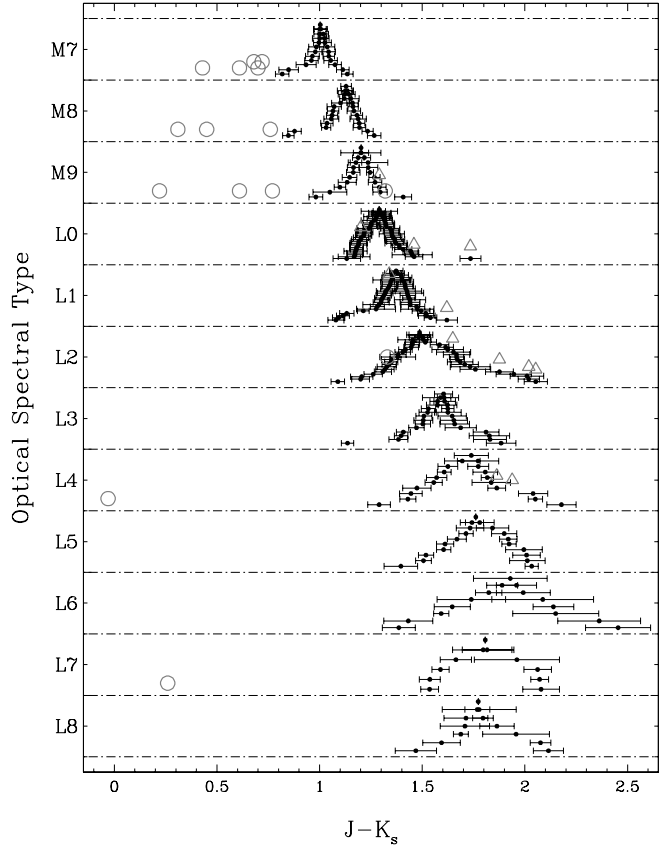}{4.7in}{0}{45}{45}{-145}{-20}
\caption{$J-K_s$ colors for a collection of M7-L8 dwarfs with optically 
determined spectral types. Each bin represents a full integral subtype: 
``M7'' includes M7 and M7.5 dwarfs, ``M8'' includes M8 and M8.5 dwarfs, etc. 
For each group of objects, the median color in the group is plotted highest 
in the bin; colors falling farther from the median are plotted progressively 
farther down the y-axis. Grey circles show the color locations of objects optically classified as subdwarfs, and grey triangles show the color locations of objects whose optical spectra show the hallmarks of lower gravity.
\label{shrimp}}
\end{figure}

With these optically typed, low-gravity late-M and L dwarfs, we can explore the effects of low gravity in the near-infrared. Figure~\ref{shrimp} shows the 2MASS $J-K_s$ colors for optically typed late-M through late-L dwarfs. Small dots with error bars are field dwarf fiducials, and objects with optical signatures of low gravity are denoted by triangles. Note that low-gravity objects tend to be redder in $J-K_s$ than normal dwarfs of the same type. An example of one low-gravity L0 in discussed in detail in \cite{kirkpatrick2006}. Features at J-band showing the hallmarks of low-gravity include retarded condensation of VO and weaker alkali lines. The redder $J-K_s$ colors are believed to be due to a weakening of the collision-induced absorption by H$_2$, which is a dominant opacity source at $H-$ and particularly $K_s-$bands.

\subsubsection{Low Metallicity}

The earliest studies illustrating effects of low metallicity on the spectra of ultra-cool dwarfs were by \cite{gizis1997}, \cite{schweitzer1999}, and \cite{lepine2003}. Lower metallicity means that fewer metal+metal molecules will be formed relative to metal+hydrogen molecules. Thus in far optical spectra, absorption by metal hydrides will be increased relative to that of metallic oxides.

How are the spectra of low-Z dwarfs affected at the transition from late-M to early-L? The hallmark of the M/L transition is the appearance of condensates, or rather the weakening of certain bands in the optical spectra as those molecules disappear into condensates. Metallicity is expected to play an important role in condensation (e.g., \citealt{lodders2002}), with low-Z meaning that fewer heavier elements will be present to form condensates in the first place.

\cite{burgasser2007} have compiled a list of known subdwarfs and extreme subdwarfs of type M7 and later. (See Burgasser et al., this volume, for in-depth discussion of spectroscopic features.) Using  optically classified objects from this list and new additions from Kirkpatrick et al.\ (in prep.) we find that they tend to have bluer $J-K_s$ colors, in some cases extremely bluer colors, than normal late-M and L dwarfs of the same optical type (circles in Figure~\ref{shrimp}). This discrepancy is believed to be largely due to collision-induced absorption by H$_2$, which is a far more dominant absorber at these wavelengths due to the decrease in abundance of metal species.

\subsubsection{Dust Content}

Despite the fact that low gravity tends to redden the near-infrared spectra and low metallicity tends to make them bluer, not all anomalously red near-infrared spectra appear to have low gravity nor do the anomalously blue ones appear to have low metallicity. 

An example of the first kind is 2MASS J22443167+2043433, whose color of $J-K_s$=2.45$\pm$0.16 makes it one of the reddest L dwarfs known. The optical spectrum of this object looks like that of a normal L6.5 dwarf, but the near-infrared spectrum shows enhanced flux at $H$- and $K_s$-bands and weak K I lines at $J$-band (\cite{mclean2003}). It is possible that this is a late-L analog to the low-gravity, early-L dwarfs discussed above. In this case the optical spectrum may be mimicking the spectrum of a normal slightly earlier L dwarf via weakening of the 7665/7699 \AA\ K I resonance doublet, which is the primary shaper of the spectrum in the far red. However, it is also possible that thicker dust clouds and veiling could produce these same effects, although the physical explanation of the thicker dust relies on further undetermined physics (e.g., higher metallicity).

An example of the second kind is 2MASS J17210390+3344160, whose $J-K_s$ color of 1.14$\pm$0.03 is at least a half magnitude bluer than a normal, optically classified L3 dwarf. One possible explanation for this object is that it has lower metallicity than a standard field L dwarf, but not as low as the subdwarfs whose optical spectra clearly identify them as low-Z. In this case the near-infrared region (because of collision induced absorption by H$_2$) may be a more sensitive indicator of metallicity than any of the optical diagnostics. An alternate explanation is a reduction of condensates in the photosphere, leading to reduced flux at $H$- and $K$-bands relative to a normal L dwarf. This could be caused either by more efficient sedimentation (the physical process for which is unknown) or by reduced metallicity. Both scenarios are discussed at greater length in \cite{cruz2007}. For additional examples of blue L dwarfs and further discussion on possible physical scenarios, see \cite{knapp2004}, \cite{chiu2006}, and \cite{cruz2007}.

\subsubsection{Unresolved Binarity}

Another factor contributing to the larger scatter in near-infrared types may be unresolved binarity. Kirkpatrick et al.\ (in prep.) synthesize composite binaries using single dwarfs with types from late-M through late-T, and show that the optical spectral composites are virtually indistinguishable from spectra of single objects. The largest deviations between the synthetic binaries and the spectra of single standards occur for late-L primaries with mid- to late-T secondaries. Looper et al.\ (in prep.) have done a similar analysis in the near-infrared. The largest deviations occur for the same set of hypothetical binaries, but the discrepancies are sometimes large enough to lead to peculiar looking spectra. See also \cite{burgasser-solo2007} for other synthetic binary analysis in the near-infrared.

\subsection{Disentangling these Effects through Classification}

One of our future challenges is finding spectral features that can reliably distinguish between these four effects. Ideally we should establish standard near-infrared sequences of normal L dwarfs and optical/near-infrared sequences of L subdwarfs and low-gravity L dwarfs. L subdwarfs may be separable into several sequences to parallel overall metal content, but the construction of such sequences is hampered by a severe lack of current examples from which to draw standards. Also, nature should preclude L subdwarfs of lowest metallicity from currently existing in the Milky Way, as brown dwarfs of such low-Z (age$\approx$10 Gyr) will have long ago cooled to type T or later.

\begin{figure}[!ht]
\plotfiddle{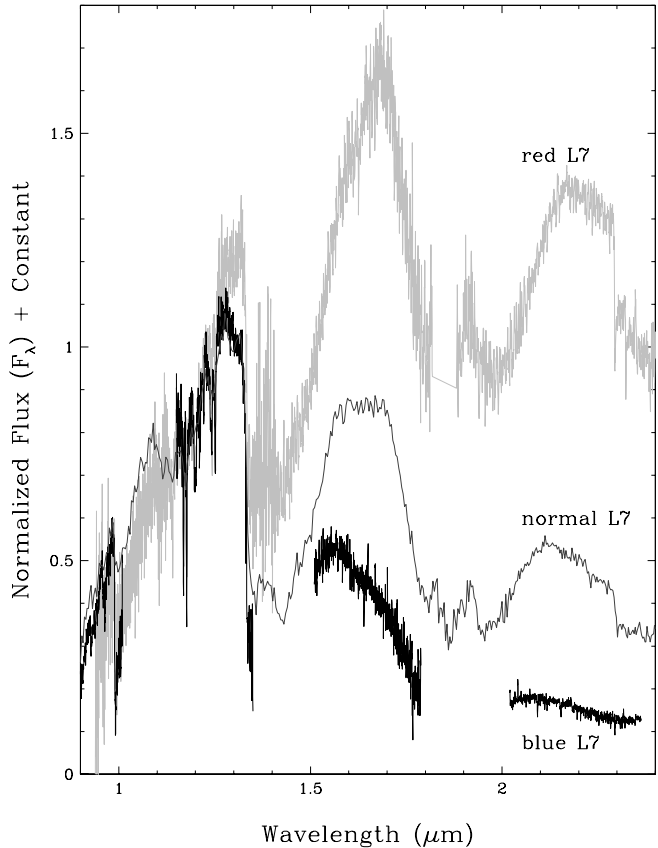}{4.7in}{0}{45}{45}{-145}{-20}
\caption{Overplot of three objects typed as $\sim$L7 in J-band. Shown are an unusually red L7 (top, light grey), a normal L7 (middle, dark grey), and a blue L7 (bottom, black). The spectra, from top to bottom, come from Looper (priv.\ comm.), Cruz (priv.\ comm.), and \cite{burgasser2003}. The blue L7 has only partial wavelength coverage in this wavelength regime. All spectra have been normalized at 1.2 $\mu$m.
\label{red_blue_L7s}}
\end{figure}

Figure~\ref{red_blue_L7s} shows what a slice at L7 might look like for three of these near-infrared spectral sequences. The top spectrum shows an L7 that is anomalously red, due either to dust effects, low-gravity, or both. The bottom spectrum shows an L7 that is anomalously blue, probably due to low-metallicity effects. In the middle is a normal L7. The extreme nature of the variations is obvious. 

Inroads have already been made into formulating these new spectral sequences and their classification schemes. \cite{gorlova2003} and \cite{mcgovern2005} have obtained near-infrared spectra of young late-M dwarfs in clusters of various ages to quantify the effects of lower gravity at these temperatures. \cite{allers2007} has devised a pair of indices at near-infrared wavelengths that can measure spectral type independent of gravity, and gravity independent of reddening, for late-M and early-L dwarfs. 

\section{Issue \#2: The L/T Transition}

Trigonometric parallax measurements have revealed another effect not predicted {\it a priori} by the models. This is the $\sim$1.3 mag brightening of $J$-band seen between late-L and mid-T. This so called ``$J$-band bump'' is illustrated in Figure~\ref{jbump}. High-resolution imaging studies of objects in the L/T transition region have revealed that the fraction of binaries with types between L7 and T3.5 is twice that seen at earlier or later types ({\citealt{burgasser-hst2006}). If all objects residing in the bump were binaries, then it might be possible to split the overluminosity of these joint systems between the components and substantially reduce the magnitude of the bump, as suggested by \cite{liu2006} and \cite{burgasser-hst2006}. 

However, not all objects occupying the bump have been successfully split. The biggest offender, 2MASS J0559$-$1404 (see Figure~\ref{jbump}),  appears single to 0$\farcs$05 with HST imaging (\citealt{burgasser-hst2003}) and to 0$\farcs$04 imaging with laser guide star adaptive optics at Keck (\citealt{gelino2005}). Even if it eventually proves to be a tight, equal-magnitude binary, this reduces its absolute $J$-band magnitude by only 0.75 mag, meaning that the $J$-band bump still has an amplitude of 0.5-0.6 mag. On-going high-resolution radial velocity measurements as well as on-going astrometric monitoring of 2MASS J0559$-$1404 will eventually reveal the truth about this object.

Furthermore, some of the L/T transition binaries that have been split show the $J$-band brightness reversal themselves. \cite{gizis2003} found the first system in which the secondary was brighter than the primary, although this observation was done not at $J$-band but at the F1042W filter ($\sim{Y}$-band) of HST. Since then three other systems have been published (\citealt{liu2006, burgasser2006, cruz2004}). A fifth binary with brightness reversals in its components has been recently reported by Looper et al.\ (in prep.) and for this system the secondary is brighter than the primary by 0.5 mag at $J$. In summary, the ``$J$-band bump'' is real with an amplitude of at least half a mag.

Regardless of the exact amplitude of the bump, we have come to learn that the relative lack of single objects at early-T is probably indicative of rapid evolution from late-L to mid-T (\citealt{burgasser-hst2006}). The bump itself my be due to a 1$\mu$m analog of the well-known 5$\mu$m ``holes'' in Jupiter's atmosphere (\citealt{gillett1969, westphal1969}). Using a scenario described in \cite{ackerman2001}, it may be the formation of holes in the cloud coverage that is responsible for the extra flux near 1$\mu$m (\citealt{burgasser-holes2002}). These holes would be relatively opacity- and cloud-free windows allowing photons from deeper, warmer layers to escape. \cite{orton1996} has shown that only $\sim$1\% of Jupiter's atmosphere contributes to the 5$\mu$m hole phenomenon, so the extent of cloud break-up may not have to be severe at the L/T boundary to account for the extra 1$\mu$m flux.

Other possible explanations, including the gravity-dependent L/T transition model of \cite{tsuji2003} and the sudden downpour model of \cite{knapp2004}, are discussed at greater length in \cite{knapp2004} and \cite{kirkpatrick2005}. See Barman (this volume) for further discussion of this and other outstanding issues being targetted by theorists.

\begin{figure}[!ht]
\plotfiddle{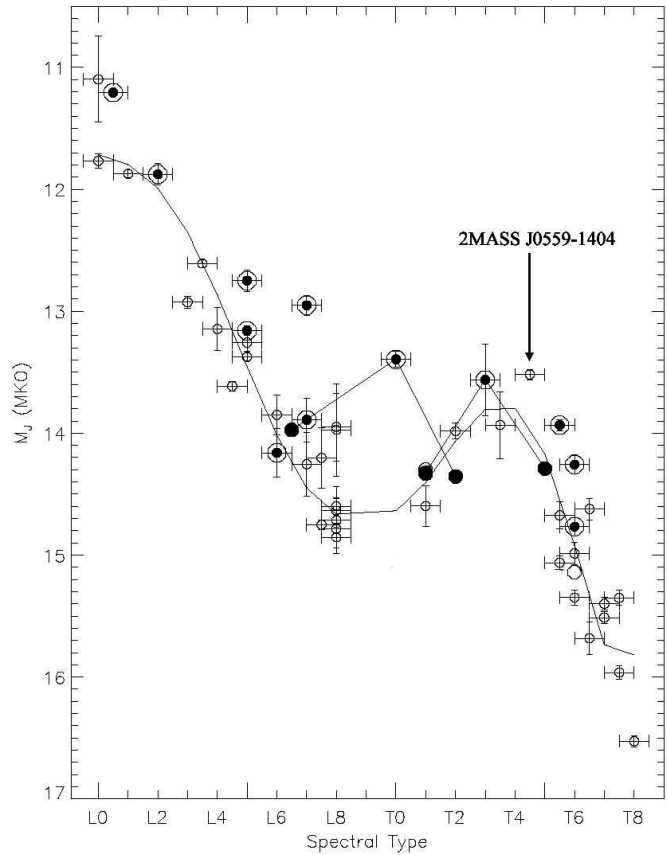}{4.7in}{0}{30}{30}{-145}{0}
\caption{Absolute $J$-band magnitude versus spectral type, adapted from \cite{burgasser-hst2006}. Optical types are shown for L dwarfs and near-infrared types for T dwarfs. Objects known to be close doubles are encircled, two of which are shown both with their joint type and magnitude as well their individual component's types and magnitudes (filled dots). The location of 2MASS J05591914$-$1404488 is singled out.
\label{jbump}}
\end{figure}

\section{Issue \#3: Finding ``Y'' Dwarfs}

Current spectroscopic classifications for brown dwarfs run as late as T8, which corresponds to $T_{\rm eff}\approx$ 750K (\citealt{vrba2004, golimowski2004}). A total of three objects, all discovered using 2MASS (\citealt{burgasser2002, tinney2005}), are typed either in the optical or near-infrared as T8 dwarfs (\citealt{burgasser_opttypes2003, burgasser2006}). The two with measured trigonometric parallaxes lie at distances of only 5.7 and 9.1 pc (\citealt{vrba2004}). This together with the fact that their limiting H-band brightnesses, 15.5$<H<$15.8, fall below the SNR=7 level of 2MASS (\citealt{skrutskie2006}), suggest that we are missing later dwarfs simply because the detectability limit of 2MASS does not enable large enough volumes to be sampled for later types.

What, then, might dwarfs cooler than T8 look like and what should they be called? If later objects are discovered with spectroscopic morphologies very similar to T dwarfs, the numbering should continue to T9, T10, or even beyond. Only when a distinct change in spectroscopic morphology -- such as the disappearance in the optical of the oxide bands at  M/L boundary or the appearance shortward of 2.5 $\mu$m of methane at the L/T boundary -- should a new class be introduced. Finding these objects, which are commonly referred to as ``Y dwarfs'' after the suggestion of \cite{kirkpatrick1999} and \cite{kirkpatrick2000}, are one of the goals of current and planned surveys (see Pinfield et al.\ and Leggett contributions, this volume).

Nature will eventually reveal what the spectra of Y dwarfs look like. In the meantime we are left to theoretical predictions to divine what the trigger at the T/Y boundary might be. Figure~\ref{ydwarfs} shows a sequence of 0.6-2.5 $\mu$m model spectra below $T_{\rm eff}$=800K from \cite{burrows2003}. Inspection of these models suggests several possible triggers, among which are (1) the appearance of ammonia in the near-infrared near $T_{\rm eff}\approx$600K, (2) the disappearance of the alkali lines near $T_{\rm eff}\approx$500K,  and (3) the end of the blueward trend of $J-K_s$ color with type near $T_{\rm eff}\approx$350K. Interestingly, the onset of water clouds near $T_{\rm eff}\approx$400-500K has no appreciable affect on the spectra.

\begin{figure}[!ht]
\plotfiddle{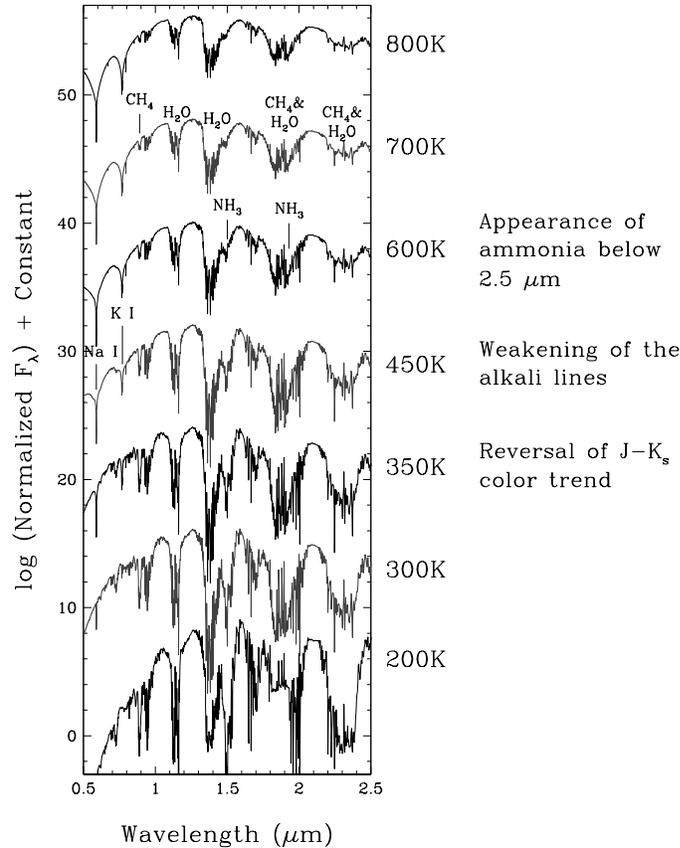}{4.7in}{0}{45}{45}{-145}{-20}
\caption{Far-optical/near-infrared spectroscopic models from \cite{burrows2003} showing solar metallicity brown dwarfs with an age of 1 Gyr and masses of 25, 20, 15, 10, 7, 5, and 2 Jupiter masses (top to bottom). Approximate effective temperatures are given along the right-hand side of the figure. Important spectral features are marked. All spectra are normalized to one at their peak flux at $J$-band, and a constant has been added to separate the spectra vertically.
\label{ydwarfs}}
\end{figure}

\acknowledgements 

The author would like to acknowledge insightful discussions with Katelyn Allers, Kelle Cruz, Dagny Looper, and Adam Burgasser during the preparation of this manuscript.


\end{document}